\documentclass[prb,twocolumn,floats,showpacs]{revtex4}
\usepackage{graphicx}
\usepackage{epsfig}
\usepackage{psfrag}
\usepackage{amssymb}
\newcommand{\gqH}{g_{\rm qH}}
\newcommand{\gM}{g_{\rm Mott}}

\newcommand{\pr}{Phys.\ Rev.\ }

\newcommand{\be}{\begin{equation}}
\newcommand{\ee}{\end{equation}}

\begin{document}

\title{Density profiles for atomic quantum Hall states}

\author{N.R. Cooper,$^1$ F.J.M.~van Lankvelt,$^{2,}$\footnote{address from Oct.~2004: Rudolf Peierls Centre for Theoretical 
Physics, 1 Keble Road, Oxford OX1 3NP, England} J.W.~Reijnders$^2$ and K.~Schoutens$^2$}
\affiliation{$^1$Cavendish Laboratory,
Madingley Road, Cambridge CB3 0HE, United Kingdom\\
$^2$Institute for Theoretical Physics, University of Amsterdam,
Valckenierstraat 65, 1018 XE~~Amsterdam, the Netherlands}



\date{1 September, 2004}

\begin{abstract}

\noindent 
We analyze density profiles for atomic quantum Hall states,
which are expected to form in systems of rotating cold atoms in the 
high-rotation limit. For a two-dimensional (single-layer)
system we predict a density landscape showing plateaus at quantized 
densities, signaling the formation of incompressible groundstates.
For a set-up with parallel, coupled layers, we predict
(i) at intermediate values of the inter-layer tunneling: a 
continuously varying density profile $\rho(z)$ across the layers, 
showing cusps at specific positions, (ii) at small values for the 
tunneling, quantum Hall-Mott phases, with individual layers at 
sharply quantized particle number, and plateaus in the density
profile $\rho(z)$.
 
\end{abstract}

\maketitle
\vspace{0.1in}

Among the fascinating developments in the field of quantum
gases is the possibility to study correlated states of matter
in a setting that is entirely different from the traditional
setting of electrons in a solid state environment. A prime
example are fractional quantum Hall (qH) states, which are expected 
when trapped atoms (bosons or fermions) are made to rotate at 
ultra-high angular momentum \cite{WGS,CW,CWG,RJ,HM,PZC,RvLSR}. 
While there has been steady progress in achieving high
rotation rates \cite{ENS-JILAexp}, the conditions for the 
actual realization of these states have not yet been met.
[See \cite{M-SDL,PPC} for alternative proposals aimed at conditions 
where atomic qH states are expected.]

The most direct experimental signature of electronic (fractional) 
qH states, the quantization of the Hall conductance,
is not easily available for realizations of such states with 
neutral atoms. It is thus important to investigate the observable
features of atomic qH states. Earlier studies have 
focused on the fractional statistics\cite{PFCZ},
loss of condensate fraction\cite{SHM1},
properties of edge excitations \cite{Ca}, and on density 
correlations in expansion images \cite{RC-ADL}. 

In this Letter, 
we work out the proposals put forward in \cite{vLRS,Co}
for the detection of atomic (fractional) qH states via 
characteristic density profiles. We present results
for a single layer, and for multi-layer configurations. 
We shall focus on the fractional qH states formed of
interacting bosons, although qualitatively similar results would be
obtained for the fractional qH states of interacting fermions.  We
remark that non-interacting fermions in a rapid rotation regime, and
subject to a slowly varying potential, will display density profiles
similar to the ones we discuss here \cite{HC}.  In that case, the
effects arise from the Landau level structure for non-interacting
fermions. All results discussed in the present work are strong
many-body interaction effects.

We first analyze a single layer situation, with rotation at or near 
the critical frequency $\omega_\perp$. The characteristic length scales 
$\ell_{\perp}$ and $\ell_\parallel$ are set by the harmonic confinement 
$\omega_\perp$ in the $x-y$ (in-plane) direction and $\omega_\parallel$
in the $z$ (out-of-plane) direction, according to
$\ell_{\perp,\parallel} = \sqrt{\hbar/(m\omega_{\perp,\parallel})}$.
The energy scale characterizing the qH states is 
$g_{\rm qH} = \frac{1}{(2\pi)^{3/2}}
             \frac{4\pi \hbar^2 a_s}{m \ell_\parallel \ell_\perp^2}$
with $a_s$ the scattering length. Assuming $a_s=5$ nm, 
$\hbar \omega_\perp \simeq 5$ nK, $\ell_\parallel= 50$ nm, $\gqH$ is 
in the order of 1~nK.
 
The state of matter formed by rapidly rotating bosonic atoms, at critical
rotation $\omega=\omega_\perp$ and in the absence of any additional 
potential, depends on the filling fraction $\nu \equiv n/n_0$ 
[$n_0 \equiv 1/\pi\ell_\perp^2$]\cite{CWG}.
For $\nu$ less than $\nu_c\sim 6$, the vortex lattice is unstable
\cite{CWG,SHM2}, and the groundstates are homogeneous quantum fluids.
Their interaction energy density $e[\nu]$ is a complicated function of 
$\nu$, containing cusps that indicate incompressible groundstates at
specific filling fractions $\nu_i$. The incompressible groundstates
have been studied by exact diagonalization studies in edge-less
geometries (sphere or torus), where it was found that they are
well-described by a variety of (fractional) qH liquids 
\cite{WGS,CW,CWG,RJ,HM,RvLSR}. These include the Laughlin
state at $\nu=1/2$, a series of states at $\nu^{(p)}= p /(p+1)$ 
that can be understood in terms of composite fermions (CF) \cite{CW},
and non-abelian states such as the Moore-Read (MR) and 
Read-Rezayi (RR) states \cite{MR-RR} at $\nu=k/2$.

Here we focus on inhomogeneous qH liquids that form in
the presence of a residual potential $V(r)$ in the rotating frame
of reference.
We consider (i) rotation $\omega$ slightly below the critical 
frequency $\omega_\perp$, leaving a residual parabolic potential 
$V_2(r)= {1 \over 2} k_2 r^2$ 
with $k_2 \propto (\omega_\perp - \omega)$, (ii) critical rotation 
$\omega=\omega_\perp$, with an additional 
confining potential, here taken to be quartic, 
$V_4(r) = {1 \over 4} k_4 r^4$. 
We shall discuss first the case in which there is a very large number of
atoms, $N\gg 1$, in which case mean density profiles can be reliably found
from a one-shot measurement of particle positions; we shall then turn to
discuss the effects of fluctuations when $N$ is not very large.

If the potential $V(r)$ varies slowly in space compared to $\ell_\perp$, 
the density distribution, $n(r)$, averaged on scales large compared 
to $\ell_\perp$, can be obtained by minimizing
\be
  E = \int d^2r \left\{ e[\nu(r)] + V(r) n(r) - \mu n(r) \right\} \ ,
\label{eq:energy}
\ee
where $e[\nu]$ is the interaction energy density arising from
the contact interactions, $\nu(r) \equiv n(r)/n_0$ and $\mu$ is the
chemical potential.  
That the energy functional (\ref{eq:energy}) is local significantly
simplifies its minimization: at each position, the density $n(r)$ is
that which minimizes $e[\nu(r)] - \mu_L n(r)$, where the local
chemical potential is $\mu_L(r) \equiv \mu - V(r)$.  For a vortex
lattice (at large $\nu$), the energy density is $e[\nu] = n_0 \nu^2 b
\, \gqH$, where $b = 1.1596$ is the Abrikosov parameter for a
triangular lattice.  The dependence of density on $\mu_L(r)$ is then
simply $n(r) \propto \mu_L(r)$, leading to a Thomas-Fermi
profile.\cite{wbp} For the qH regime, $\nu< \nu_c$, there are cusps in
the energy function $e[\nu]$, which give rise to a step-like
dependence of $n$ on $\mu_L$; this causes plateaus in the spatial
density distribution.

As a simple example, we consider a situation where liquids at 
$\nu = 1/2$ and $\nu=2/3$ make up the lowest energy
configuration. The energy per unit area takes the values
$e[\nu=1/2]=0$, and $e[\nu=2/3] = (2/3) n_0 \epsilon_2$
(with $\epsilon_2$ a number of order $\gqH$, see \cite{RJ}). 
We find $\nu = 0$ for $\mu < 0$, $\nu = 1/2$ for 
$0< \mu < 4 \epsilon_2$, and $\nu = 2/3$ for $\mu > 4\epsilon_2$.

Let us take a harmonic confining potential $V(r) = {1 \over 2} k_2 r^2$, 
such that
$ \mu_L(r) = \mu - {1 \over 2} k_2 r^2$. The $\nu=1/2$ state forms a disc that 
extends out to $\mu_L=0$, {\it i.e.} to a radius
$r_1 = \sqrt{2\mu/k_2}$. There is a 
critical value of chemical potential, $\mu_c = 4 \epsilon_2$, at which 
the $\nu=2/3$ state will first appear in the center of the trap 
(where $\mu_L$ is maximum). At $\mu=\mu_c$, the number of atoms in the 
$\nu=1/2$ disc is $N_c =  1/2 \, n_0 \pi r_1^2  = 
2 \epsilon_2 / \lambda_2$, with $\lambda_2 = \frac{1}{2} k_2 \ell_\perp^2$. 
Above this critical value, the 
inner disc at $\nu=2/3$ has a radius $r_2=\sqrt{2(\mu-\mu_c)/k_2}$. 
Expressed in terms of $N$ and $N_c$, the locations of the two steps 
are $r_2/\ell_\perp = \sqrt{3(N-N_c)/2}$,
$r_1/\ell_\perp = \sqrt{(3N + N_c)/2}$.
From the dependence $e[\nu]$ obtained from exact diagonalizations\cite{CWG} 
it can be inferred that intermediate
densities other than $\nu=0, 1/2, 2/3$ will not appear in the
density profile. 

Repeating this analysis in a quartic potential 
$V_4(r)={1 \over 4} k_4 r^4$ leads to a critical $N$ of
$N_c = \sqrt{\epsilon_2 / \lambda_4}$,
with $\lambda_4 = {1 \over 4}k_4 \ell_\perp^2$, with inner and
outer edges at $(r_2/\ell_\perp)^2 = {3 \over 4}(\sqrt{9N^2-8N_c^2}-N)$,
$(r_1/\ell_\perp)^2 = {9 \over 4} N - {1 \over 4} \sqrt{9N^2-8N_c^2}$.

\begin{figure}
\epsfig{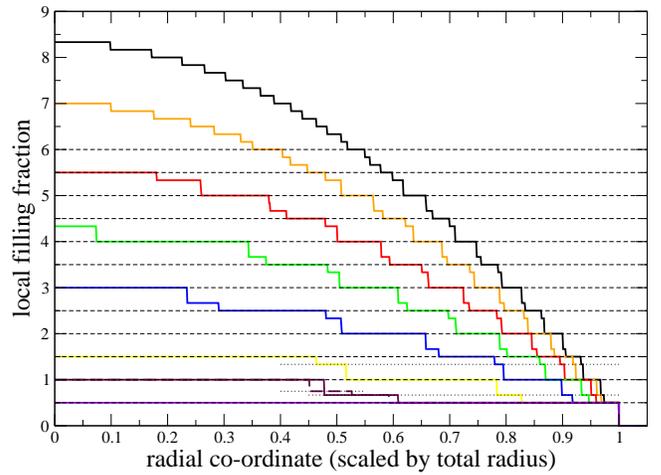} \caption{Radial density profiles in a
single layer with harmonic confinement, for a series of filling
fractions at the center of the trap. The solid and dashed lines
correspond to energy functions $e[\nu]$ obtained from numerics using
$N_V=6$ and $N_V=12$ states, respectively.}
\label{fig:steps}
\end{figure}

Including in the analysis more quantum liquids, at filling 
$\nu_1<\nu_2<...$, leads to a sequence of critical values $N=N_c^{(k)}$, 
marking the onset of the formation of a central region of a qH 
fluid at $\nu=\nu_k$. To determine the corresponding density profiles
we use the interaction energy function $e[\nu]$ calculated by exact 
diagonalization studies on a torus \cite{CWG}. In the graphs presented in 
Fig.~\ref{fig:steps}, a 
confining potential $V(r) = \frac{1}{2} k_2 r^2$ is assumed. 
The energy functions used are for
a system size of $N_V= 6$ single-particle states on a torus with aspect ratio
$\sqrt{3}/2$ (solid lines) [with numerical results up
to $\nu=\frac{25}{3}$], and for $N_V = 12$ single-particle states on a torus
with aspect ratio $0.3$ (dashed lines) [up to $\nu=\frac{7}{6}$].
For filling fractions larger than $\nu_c \sim 6$, the groundstate is a
compressible vortex lattice \cite{CWG}. In this regime, the density
profile (averaged on a lengthscale large compared to $\ell_\perp$) is
an inverted parabola\cite{wbp}. The steps seen in Fig. 1 in this
regime are an artifact of the finite-size numerics (on a finite system
with $N_V=6$ the minimum change of filling fraction is $\Delta \nu =
1/6$).
For filling fractions less than $\nu_c \sim 6$, plateaus appear at the
filling fractions of incompressible qH fluids, including the sequence of MR
and RR states at $\nu=k/2$ [horizontal dashed lines] as well as some
of the CF sequence $\nu = p/(p+1)$ with $p=2, 3, -4$ [dotted
lines]. This plateau structure is a direct consequence of the
incompressibility of the fractional quantum Hall liquids. This
structure could be observed in the density distribution of a
single-layer system following expansion\cite{RC-ADL}.



We have tested the above picture against exact
diagonalization results for small systems. However, at the small sizes 
that can be handled in exact diagonalization (up to $N=12$ particles 
on a disc), the density profiles seem to be dominated by edge effects, 
indicating that the system sizes do not exceed the correlation lengths
in the expected qH liquids \cite{CvLRS}.

We next investigate the situation where an optical lattice in
the $z$ direction is imposed on a cloud of
atoms rotating around the $z$-axis.\footnote{The authors were 
first made aware of this proposal by E.~Cornell.} If the optical
potential is sufficiently deep, this will define a stack of
parallel planes with (weak) tunneling $t$ between the planes.
The lengthscale $\ell_\parallel$ is now determined 
by the confinement frequency in a single layer of the optical lattice.
We assume that there is an overall quadratic confinement in the 
$z$-direction, giving rise to a chemical potential 
$\mu(z) = \mu_0 - \mu_2 (z/d)^2$ (with $d$ the distance between 
the layers), which will induce a slow modulation in the number 
of particles per layer.
The idea is that parameters can be chosen such that, while 
the original cloud is in a mean field regime ($\nu>\nu_c$), displaying 
a vortex lattice, the individual layers defined by the 
optical lattice can be in a quantum regime, so that the entire 
configuration becomes a stack of weakly coupled quantum 
liquids.
With $N_L$ layers, a total filling fraction $\nu$ (for the entire
cloud) gives a filling fraction per
layer of $\nu^\prime=\nu/N_L$. If $\nu^\prime < \nu_c$, the vortex 
lattice will melt if the inter-layer tunneling energy $t$ is made 
smaller than a critical value $t_{c_1}$. We have generalized the 
analysis of \cite{SHM2} to determine
the value of $t_{c_1}$ (see \cite{CvLRS} for a detailed account).
For $N=5000$ particles, $N_V=100$ vortices and a number 
of layers $N_L=50$, our estimate is $t_{c_1} \sim 0.1 \, \gqH$.

Assuming $N/N_L\gg 1$ and $t\ll \gqH$ we have the following physical 
picture. Within each layer, we have a density landscape built out of
incompressible qH liquids each of the form shown in Fig. 1. This landscape varies slowly 
from layer to layer. 
The gap for bulk excitations over each 
qH liquid is of order $\gqH$ \cite{CWG,RJ} and since 
$t\ll \gqH$ there will be no `bulk-to-bulk' tunneling events between 
the layers. However, the energy scale for processes where atoms tunnel 
from the edge of a qH fluid in one layer $i$ to the 
corresponding edge in an adjacent layer $i\pm 1$, is much lower,
of order $\gM$. This scale is defined as $\gM \simeq \lambda_2$ for 
quadratic confinement and $\gM \simeq \lambda_4 N_i$ for the quartic 
case, with $N_i$ the number of particles in the qH 
liquid in layer $i$. In terms of $N_0= \gqH/\lambda_2$ 
($N_0=\sqrt{\gqH /\lambda_4}$) we have $\gM\simeq \gqH/N_0$ 
($\gM\simeq \gqH N_i/N_0^2$) for the quadratic (quartic) case.
if the tunneling $t$ is large on the scale of $\gM$, inter-layer 
tunneling events will establish a continuously 
varying density profile $\rho(z)=\langle N_i \rangle$ 
across the layers (where $N_i$ is now the total number of particles 
in layer $i$). This profile could be measured from an image of 
the confined condensate, provided the period of the lattice,
$d$, is made larger than the imaging resolution (perhaps by forming
the optical potential from running waves at a shallow angle). 
\footnote{The authors learned of this proposal from J.~Dalibard.} 

We first assume quadratic in-plane confinement. If
the number of particles per layer stays below 
$N^{(2)}_c = 2 \epsilon_2 / \lambda_2$, 
the qH liquids in the layers will be Laughlin
liquids at $\nu=1/2$. Minimizing the total energy w.r.t. 
each of the $N_i$, we obtain a parabolic TF density profile
\be
\rho(z) = \frac{1}{d^2} {\mu_2 \over 2\lambda_2}
          (z_1^2-z^2) \ ,
\ee
with 
$z_1^3 = d^3 {3N \over 2}{\lambda_2 \over \mu_2}$. If
$N$, $\mu_2$ are such that some of the $N_i$ exceed the critical 
values $N^{(k)}_c$, there will be additional stable qH 
liquids in these layers. The corresponding profile $\rho(z)$
exhibits cusps at the positions $\pm z_k$ where $N_i$ goes 
through $N^{(k)}_c$. If the central layers near $z=0$ support 
$\nu=1/2$ and $\nu=2/3$ liquids, we have 
\begin{eqnarray}
\rho(z) & = & \frac{1}{d^2} {\mu_2 \over 2\lambda_2} (z_1^2-z^2)
	      \quad \quad {\rm for} \quad z_2<|z|<z_1
\nonumber \\
        & = &  N_c^{(2)} + 
            \frac{1}{2d^2} {2 \mu_2 \over 3 \lambda_2}(z_2^2-z^2) 
	    \quad {\rm for} \quad |z|<z_2
\end{eqnarray}
with $z_1^2-z_2^2= 2d^2 N_c {\lambda_2 \over \mu_2}$
and $z_1^3+z_2^3/3= d^3 {3N \over 2}{\lambda_2 \over \mu_2}$.
This profile shows a (weak) cusp at $|z|=z_2$, with the slope 
${\partial \over \partial z}\rho(z)$ changing by a factor of
4/3. 

\begin{figure}
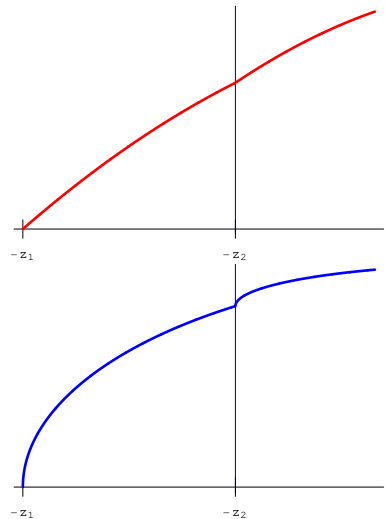

\epsfig{file=profilequadra.epsi,width=55mm} 
\epsfig{file=profilequartic.epsi,width=55mm} 
\caption{Density profiles $\rho(z)$ for quadratic (top) 
and quartic (bottom) in-plane confinement. The layers
at $-z_1<z<-z_2$ have the $\nu_1=1/2$ liquid only; the
cusps at $z=-z_2$ mark the onset of a second liquid
with $\nu_2=2/3$ in the layers.} 
\label{fig:profiles}
\end{figure}

\begin{figure}
\epsfig{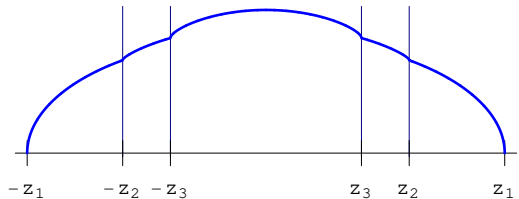} 
\caption{Schematic density profile $\rho(z)$, for quartic in-plane 
confinement and qH liquids at $\nu_1=1/2$, $\nu_2=2/3$, $\nu_3=1$.}
\label{fig:fullprofile}
\end{figure}

The nature of the cusps at $z=z_k$ depends on the in-plane
potential $V(r)$. In Fig.~\ref{fig:profiles} we show 
the cusp at $z=-z_2$ for two choices of $V(r)$. Clearly, a steeper 
confinement in the layers leads to a more pronounced cusp in the 
profile $\rho(z)$. [A square well cut-off in $V(r)$ can 
produce a flat plateau in $\rho(r)$.] For quartic confinement, 
the profile at $N<N_c^{(2)}$
is a semi-circle, and the cusps at $|z|=z_k$ have square-root
singularities. In fact, $\rho(z)$ is of the general form $\rho(z) 
= \sum_k a_k \sqrt{z_k^2-z^2}$. Fig.~\ref{fig:fullprofile}
shows $\rho(z)$ for quartic confinement, with the layers supporting 
quantum liquids at $\nu_1=1/2$, $\nu_2=2/3$, $\nu_3=1$.
(We note in passing that similar considerations lead to the conclusion
that the density distribution following an {\it expansion} of a
multi-layered system in an otherwise quadratically-confined trap will
have a dependence on the radial co-ordinate $r$ that is of the same
qualitative form as Fig.~\ref{fig:fullprofile} is of the axial 
co-ordinate $z$.)

We now turn to consider the situation when $N$ is not very much 
larger than 1. We begin by estimating the fluctuations. A 
deviation $\delta N_i$ from the mean field profile costs 
an energy of order $\gM (\delta N_i)^2$, giving 
\be
\delta N_i \simeq \sqrt{N_0 {t \over \gqH}} \ ,
\quad
\delta N_i \simeq \sqrt{{N^2_0 \over N_i} {t \over \gqH}}
\label{eq:deltaNi}
\ee
for the quadratic and quartic cases.
In addition there will be intra-layer fluctuations, 
for example when a particle at radius $r_i$ is promoted
to an unoccupied orbital at $r_i + \delta r_i$. 
Equating the corresponding energy to $t$ leads to
\be
{\langle \delta r_i \rangle \over r_i} 
\simeq {N_0 \over N_i}{t \over \gqH} \ ,
\quad
{\langle \delta r_i \rangle \over r_i} 
\simeq \left( {N_0 \over N_i} \right)^2 {t \over \gqH}
\ee
for the quadratic and quartic cases. This shows that if $t\ll \gqH$
and $N_i/N_0$ is not too small, the steps in the in-layer density 
profiles are well-defined on the scale of the overall radius. 

A one shot experiment will obviously produce integer values
for all $N_i$. Based on the above, we predict that these numbers
will follow our continuous curves for $\rho(z)$ rather closely,
with fluctuations as specified in Eq.~(\ref{eq:deltaNi}).
Gradually lowering $t$ (for example, by turning on the potential 
slowly enough) will lead to values $N_i$ that are nearest integers to
the corresponding values of $\rho(z)$.

\begin{figure}
\epsfig{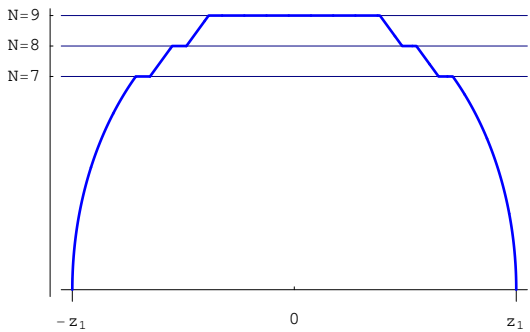} 
\caption{Schematic density-profile $\rho(z)$, for layers supporting
Laughlin liquids in quartic confinement. The horizontal segments 
indicate the qH-Mott phases, with sharply quantized particle number 
per layer, while the remaining parts correspond to `conducting' phases, 
with fluctuating $N_i$.}
\label{fig:qHMottprofile}
\end{figure}

The estimates in Eq.~(\ref{eq:deltaNi}) 
indicate that $\delta N_i \simeq 1$ if $t$ is lowered
to a value of order $\gM$. Based on the analogy with 
single atoms in an optical lattice, we anticipate a 
Mott phase where the numbers $N_i$ are sharply quantized.
We have analyzed an effective Bose-Hubbard model, with
layers supporting a $\nu=1/2$ state in quadratic confinement.
In a mean-field treatment we find, for $\mu$ satisfying 
$2\lambda_2(N-1)<\mu<2\lambda_2N$, the following critical 
$t$ for entering the $N$-particle qH-Mott phase
\be
t_{c_2}(N) = -\lambda_2
  [N - {\mu \over 2 \lambda_2}][(N-1)- {\mu \over 2\lambda_2}] \ .
\ee
This has a maximum of $\lambda_2/4$, which is of order $\gM$, 
for all $N$. For confinement `steeper than' quadratic, the 
maximal $t_{c_2}(N)$ grows with $N$, consistent with 
$\gM\simeq \lambda_4 N$ for the quartic case.

Scanning the sample from layer to layer, there will then be segments 
showing qH-Mott phases, and `conducting' regions 
in between, see Fig.~\ref{fig:qHMottprofile}. A measurement of
the particle number per layer in an $N$-particle qH-Mott 
phase will give $N_i=N$ with no fluctuations. The same measurement 
in a `conducting' phase will produce integer numbers as well, but
these will show fluctuations.

We thank J.~Dalibard and V.~Schweikhard for illuminating discussions.
FJMvL thanks Brookhaven Nat. Lab. for hospitality. This research was 
supported by EPSRC grant GR/S61263/01 (NRC) and by NWO and FOM of the 
Netherlands (FJMvL, JWR, KS).


\begin{thebibliography}{99}

\bibitem{WGS}
N.K.~Wilkin, J.M.F.~Gunn, R.A.~Smith, 
\prl {\bf 80}, 2265 (1998);
N.K. Wilkin, J.M.F. Gunn, 
\prl {\bf 84}, 6 (2000).

\bibitem{CW}
N.R. Cooper, N.K. Wilkin, 
\pr {\bf 60}, R16279 (1999).

\bibitem{CWG}
N.R. Cooper, N.K. Wilkin, J.M.F. Gunn, 
\prl {\bf 87}, 120405 (2001).

\bibitem{RJ}
N.~Regnault, Th.~Jolicoeur, \prl {\bf 91}, 030402 (2003);
cond-mat/0404093; cond-mat/0406013.

\bibitem{HM}
T.-L.~Ho, E.J.~Mueller, \prl {\bf 89}, 050401 (2002).

\bibitem{PZC}
B. Paredes, P. Zoller, J.I. Cirac, \pra {\bf 66}, 033609 (2002).

\bibitem{RvLSR}
J.W.~Reijnders {\it et al.},
\prl {\bf 89}, 120401 (2002); 
\pra {\bf 69}, 023612 (2003).

\bibitem{ENS-JILAexp}
V. Bretin {\it et al.},
\prl {\bf 92}, 050403 (2004); 
V.~Schweikhard {\it et al.}, \prl {\bf 92}, 040404 (2004).

\bibitem{M-SDL}
E.J.~Mueller, cond-mat/0404306;
A.~S{\o}rensen, E.~Demler, M.~Lukin, cond-mat/0405079;

\bibitem{PPC}
M.~Popp, B.~Paredes, J.I.~Cirac, cond-mat/0405195.

\bibitem{PFCZ}
B. Paredes {\it et al.}, 
\prl {\bf 87}, 010402 (2001).

\bibitem{SHM1}
J. Sinova, C.B. Hanna, A.H. MacDonald, \prl {\bf 90}, 120401
(2003).

\bibitem{Ca}
M.A.~Cazalilla, \pra {\bf 67}, 063613 (2003); \\
M.A.~Cazalilla, N. Barber\'an, cond-mat/0408540.

\bibitem{RC-ADL}
N. Read, N.R. Cooper,
\pra {\bf 68}, 035601 (2003);
E. Altman, E. Demler, M.D. Lukin, 
cond-mat/03096226.

\bibitem{vLRS}
F.J.M.~van Lankvelt, J.W.~Reijnders, K.~Schoutens,
presentation S3.003 at 2004 APS March Meeting (Montreal).

\bibitem{Co}
N.R.~Cooper, presentation at the conference
`Quantum Gases' KITP Santa Barbara (May 2004).

\bibitem{HC}
T.-L. Ho, C.V. Ciobanu, 
\prl {\bf 85}, 4648 (2000).

\bibitem{SHM2}
J.~Sinova, C.B.~Hanna, A.H.~MacDonald,
\prl {\bf 89}, 030403 (2002).

\bibitem{MR-RR}
G.~Moore, N.~Read, 
Nucl.\ Phys.\ {\bf B 360}, 362 (1991);
N.~Read, E.~Rezayi, \prb {\bf 59}, 8084 (1999).

\bibitem{wbp}
G.~Watanabe, G.~Baym, C.J.~Pethick cond-mat/0403470;
N.R.~Cooper, S.~Komineas, N.~Read cond-mat/0404112.

\bibitem{CvLRS}
N.R.~Cooper, F.J.M.~van Lankvelt, J.W.~Reijn\-ders and
K.~Schoutens, manuscript in preparation.

\end{thebibliography}
\end{document}